\documentclass[aps,pra,onecolumn,showpacs]{revtex4}
\usepackage{epsfig}
\usepackage{hyperref}

\begin{document}
\draft 
\def\il{I_{low}} 
\def\iu{I_{up}} 
\def\eeq{\end{equation}}
\def\ie{i.e.}  
\def\etal{{\it et al. }}  
\def\prb{Phys. Rev. B}
\def\pra{Phys. Rev. A} 
\def\prl{Phys. Rev. Lett. }
\def\pla{Phys. Lett. A } 
\def\pb{Physica B}
\def\ajp{Am. J. Phys. }  
\def\mpl{Mod. Phys. Lett. B} 
\def\ijmp{Int. J. Mod. Phys. B} 
\def\ijp{Ind. J. Phys. }
\def\ijpap{Ind. J. Pure Appl. Phys. }
\def\ibmjrd{IBM J. Res. Dev. }
\def\pjp{Pramana J. Phys.}
\def\ltp{J. of Low Temp. Phys.}

\title{Features in Evanescent Aharonov-Bohm interferometry.}
\author{Colin Benjamin}
\email{colin@iopb.res.in}
\affiliation{Institute of Physics, Sachivalaya Marg, Bhubaneswar 751 005,
  Orissa, India}
\author{A. M. Jayannavar}
\email{jayan@iopb.res.in}
\affiliation{Institute of Physics, Sachivalaya Marg, Bhubaneswar 751 005,
  Orissa, India}

\date{\today}

\begin{abstract}
  
  In this work we analyze an Aharonov-Bohm interferometer in the
  tunneling regime. In this regime, current magnification
  effect which arises in presence of transport currents is absent. A
  slight modification in the form of a quantum well incorporated in
  one of the arms leads to revival of current magnification.
  Systematics in magneto-conductance oscillations are observed in this
  evanescent wave geometry. In this framework we also see absence of
  Fano lineshapes in transmission resonances but once again one can
  recover these if the direct path supports propagating modes.

\end{abstract}


\maketitle 
\section{Introduction}

The Aharonov-Bohm(AB) interferometer is one of the most exciting
topics in mesoscopic physics research, not only because of the
fundamental physical insight into questions of quantum
non-locality\cite{steinberg} but also for being the basis of many
novel phenomena at mesoscopic scales. Some of which are, persistent
currents in normal metal rings\cite{webb,levy}, Aharonov-Bohm
oscillations\cite{gia}, Fano resonances\cite{kobay}, and current
magnification\cite{colin_cm,deo_stub}. In this work we analyze the local
currents and conductance in this interferometer but for evanescent
mode propagation. Evanescent means the incident energy of charge
carrier's in the leads attached to the interferometer is less than the
potential $V$ which characterizes the ring throughout (see FIG.~1, for 
a schematic representation of our system).
Thus when an electron with energy $E<V$ impinges on the ring it has to
tunnel out of the ring. In certain situations we also consider a
quantum well inserted in one of the arms of the interferometer. Only
in the well and the connecting leads does the electron propagate with
a real wave-vector $k=\sqrt{E}$, elsewhere it propagates with a
complex wave-vector $k=i\kappa=i\sqrt{V-E}$.  In such situation the
contribution to conductance arises simultaneously from two
non-classical effects namely, Aharonov-Bohm effect and quantum
tunelling\cite{switch}. Such situation can arise when the transverse
width of the ring is much less than the connecting ideal wires. In
this case due to the higher zero point energy arising from transverse
confinement, the fundamental sub-band minima in the ring will be at
higher energy than the value of few sub-band minima in the ideal
connecting wires. Now a situation can arise where propagating modes in
the wire have energy less than the minimum propagating sub-band energy
of the ring system. Thus the electron propagating in a lower sub-band
of the ideal wire feels a barrier to it's motion (experiences an
effective potential barrier $V$, arising solely because of the mismatch
of the zero-point energies) and tunnels across the system via
evanescent mode propagation. For simplicity we have restricted our
calculation to single channel case.  We are interested only in
coherent transport, i.e., absence of dephasing. Using the wave guide
method we have solved for all the scattering coefficients and wave
amplitudes in the ring, required to calculate the local currents in
the arms of the interferometer. For details we refer
to Refs.\cite{xia,colin_cm,deo_stub,coupled}. Expressions for the scattering
amplitudes can be obtained readily with the use of a simple MAPLE
program\cite{maple}. However analytical forms of these expressions are 
too lengthy. Therefore we analyse our results by suitable plots
generated from the obtained MAPLE expressions.

\begin{figure}[b]
\protect\centerline{\epsfxsize=3.0in\epsfbox{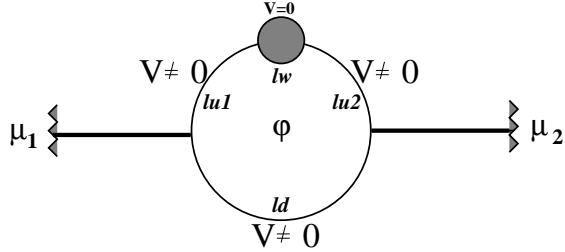}}
\caption{The Aharonov-Bohm interferometer in the evanescent regime
  with a quantum well. The strength of barrier potential is $V$, while
  arm lengths are denoted in the figure.}
\end{figure}

Herein we particularly analyze three configurations of the
Aharonov-Bohm interferometer and see how the conductance and local
currents in the two arms respond to the configurational
variations. The first configuration is an AB ring at potential $V$
attached to two leads at zero potential. In the second configuration
we consider an AB ring at potential $V$ throughout apart from the
quantum well which is at zero potential, and in the final
configuration we consider an AB ring with potential $V$ only in the
upper arm in addition to the quantum  well but the lower arm is set
free($V=0$). In particular, we show that the first configuration does
not support current magnification, while the second configuration
supports it.  Further the second configuration is marked by absence of 
Fano resonances as opposed to the third configuration. The magneto
conductance shows systematic behavior as a function of flux in the
evanescent regime. The detailed analysis of these phenomena is given
in the following sections.

\section{Evanescent Aharonov-Bohm interferometry}

\begin{figure*}
\protect\centerline{\epsfxsize=5.5in\epsfbox{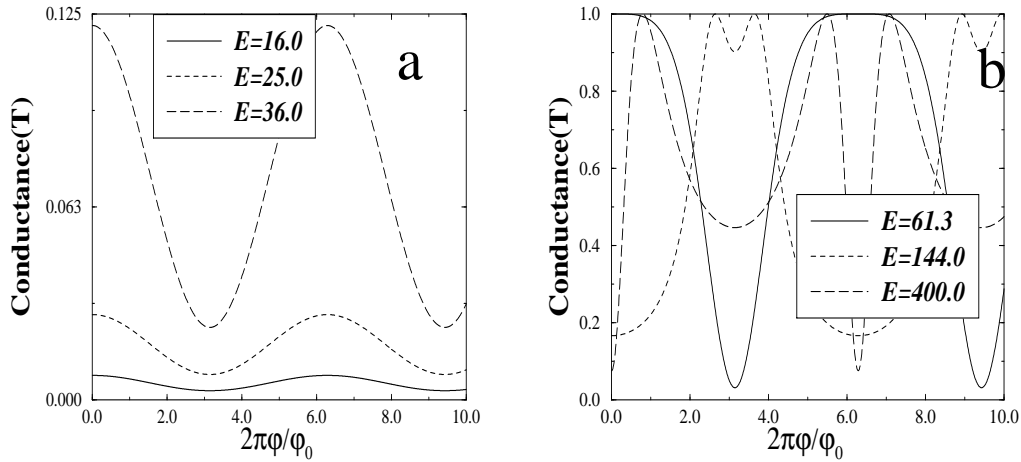}}
\caption{(a) Transport in evanescent regime ($E < V$) for different Fermi
  energies. Systematics in magnetoconductance slopes. The lengths are-
  upper arm $=0.375$, lower arm$=0.625$, Barrier Potential $=49.0$
  (all in dimensional units). (b) Transport in propagating regime($E >
  V$) for different Fermi energies. Unsystematic nature of
  magnetoconductance slopes. Same physical parameters as in
  FIG.~2(a).}
\end{figure*}

In the first configuration, we compare and contrast the cases of
transport in the pure evanescent and pure propagating mode regime,
with emphasis on it's conductance and local currents. In these cases
the well in upper arm is absent. In FIG.~2(a), we plot the conductance
in dimensionless form which is just the transmission $G=(2e^2/h)T$, as
a function of the magnetic flux for different Fermi energies below the
barrier potential, i.e., in the evanescent regime. The total length of
the ring $l (=lu1+lw+lu2+ld)$ is taken to be unity and all other
physical parameters of the ring are scaled with respect to this (we
have set $\hbar=2m=1$ throughout), for example energy in dimensionless
form is $El^{2}\equiv E$, and barrier potential $Vl^{2}\equiv V$. As
expected the conductance is periodic in flux with period $\phi_0$, the
fundamental flux quantum $\frac{hc}{e}$. We observe that the small
field differential magneto conductance is always negative. In
FIG.~2(b), we plot the conductance as a function of the magnetic flux
for energies above the barrier potential, i.e., in the propagating
regime.  We see that the small field differential magneto-conductance
(MC) may be positive/negative. From these two figures we observe the
systematic behavior of magneto-conductance slopes for pure evanescent
wave transport in contrast to the unsystematic nature of
magneto-conductance slopes for pure propagating wave
transport. The systematics in transport refer to the slope of the
magneto-conductance for small fields (i.e., the small field
differential magneto-conductance), which for pure evanescent mode
transport is always negative irrespective of the position of
impurities or system parameters. Unsystematic behavior refers to this
slope being very sensitive to system parameters or impurities, i.e.,
whether it is positive or negative cannot be predicted a priori\cite{switch}.

\begin{figure*}
\protect\centerline{\epsfxsize=5.5in\epsfbox{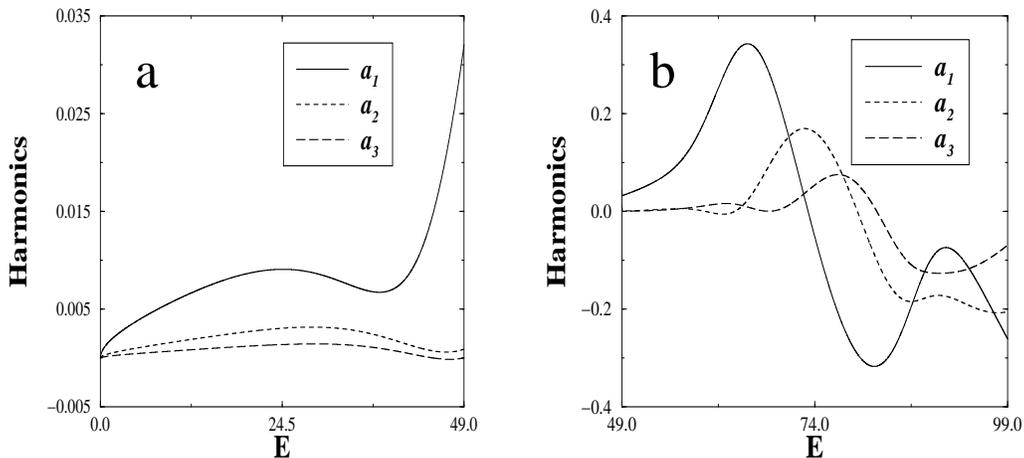}}
\caption{(a) Complete evanescent transport. Plotted are the harmonics.
  Same physical parameters as in FIG.~2(a). (b) Complete propagating
  mode transport. Plotted are the harmonics. Same physical parameters
  as in FIG.~2.(a).}
\end{figure*}

\begin{figure*}
\protect\centerline{\epsfxsize=5.5in\epsfbox{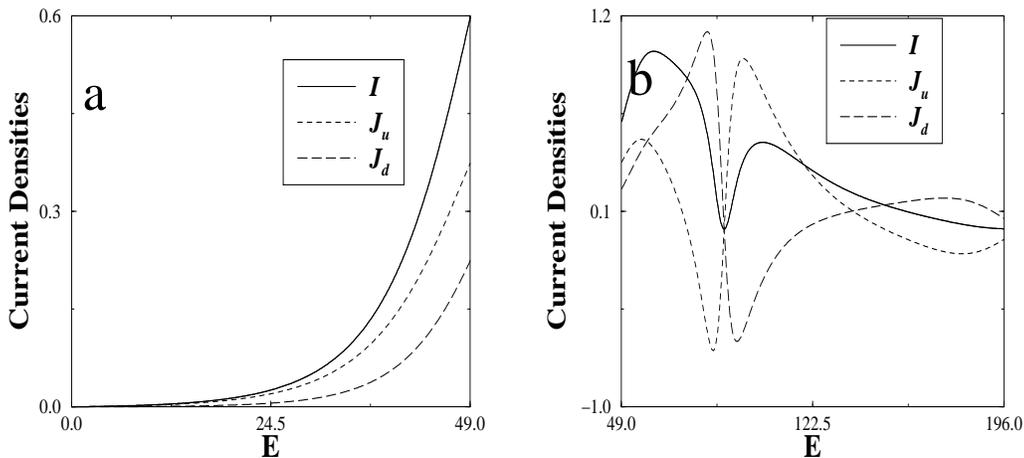}}
\caption{(a)Current densities in the evanescent regime for  zero field
  and increasing Energy. No current magnification. Same physical
  parameters as in FIG.~2(a).  (b) Current densities for the
  propagating regime for zero field and increasing Energy. Current
  magnification evident.  Same physical parameters as in FIG.~2.(a).}
\end{figure*}

The reason we observe the systematic nature of magneto-conductance
slopes has to do with the harmonics. The harmonics are calculated as
follows\cite{colin_de}-

\begin{eqnarray}
a_{n}=\frac{1}{\pi}\int_{0}^{2\pi} T cos(n\phi) d\phi
\end{eqnarray}

Herein, $T$ represents the total transmittance and $\phi$ the enclosed
flux.  The magnitude of $nth$ order harmonic corresponds to the
contribution from electronic paths which encircle the flux $n$ times.
In the case of evanescent mode propagation the first harmonic always
dominates the rest, as is seen in FIG.~3(a).  Owing to the decaying
nature of evanescent modes, the higher harmonics are weighted by a
factor $e^{-n\kappa l}$.  Thus higher harmonics, in the evanescent
regime which have to encircle the ring many more times naturally decay
faster(exponentially), therefore first harmonic completely dominates
over the rest, and thus we see systematic behavior in the conductance
and hence small field magneto-conductance slope is always negative.
This definite feature (zero flux conductance is always greater than
the conductance for increasing values of flux upto $\phi_{0}/2$, i.e.,
the small field regime) may be utilized in some device application
which has been suggested earlier\cite{switch}. For propagating waves
any harmonic can dominate over the rest (i.e., amplitude of any higher
harmonic can dominate the lower one). This is amply clear from
FIG.~3(b), and thus we see unsystematic nature of slopes for small
field differential magneto-conductance. In the case of propagating
modes the nature of the slopes is very sensitive to system parameters
and defects if present\cite{switch}.

Still considering transport in the Aharonov-Bohm interferometer
without the well, we plot in FIG's.~4(a) and 4(b), the output current
density in dimensionless form)\cite{colin_cm} as well as the current
densities in upper $J_u$ and lower arms $J_d$ as a function of the
incident Fermi energy for both evanescent(FIG.~4(a)) and
propagating(FIG.~4(b)) modes of transport.  In FIG.~4(a), we observe
that the currents in the upper and lower arms are individually less
than the output current, thus indicating the absence of current
magnification. Now what is current magnification? In the AB
interferometer, we see that the dimensionless output current density I
(=T the transmission probability) around a small Fermi energy
interval, flows through the system when voltage bias
($\mu_{1}-\mu_{2}=eV$) is applied.  The upper and lower arms of the
ring are of different lengths such that current densities $J_{u}$ and
$J_{d}$ flow in these with $J_{u} \neq J_{d}$, but $I=J_{u}+J_{d}$,
i.e., obeying Kirchoff's law of current conservation.  In particular
we see that for some particular ranges of Fermi energy $J_{u}$ or
$J_{d}$ can be much larger than $I$. Current conservation thus
dictates $J_{d}$ or $J_{u}$ to be negative\cite{colin_cm}. This
property that current in one of the arms is larger than the transport
current is referred to as current magnification effect\cite{deo_stub}.
It is an effect without any classical analog in a DC
circuit\cite{deo_stub}.  The negative current flowing in one arm can
be interpreted as a circulating current that flows continually in the
ring. The magnitude of the negative current in one of the arms flowing
against the direction of applied current is taken to be that of the
circulating current. When negative current flows in the upper arm the
circulating current direction is taken to be anti-clockwise (or
negative) and when it flows in the lower arm the circulating current
direction is taken to be clockwise (or positive)\cite{colin_cm}. It
has been suggested that due to this effect one can observe large
orbital magnetic moment of a ring in absence of magnetic field,
however, in presence of transport currents (in non-equilibrium state).
This effect has been extended to thermal currents\cite{mosk} and to
spin currents in the presence of Aharonov-Casher flux\cite{choi}.  One
can clearly see that current magnification effect is absent in the
evanescent regime, as in FIG.~4(a), while being present for the
propagating regime, as in FIG.~4(b). In particular from FIG.~4(b), one
can see that for Fermi energy range $80.0.0<E<100.0$, $J_{d}>I$ and
$J_u$ is negative, while $J_{u}>I$ for $100.0<E<121.0$ and $J_d$ is
negative. Preferentially one observes current magnification around the
anti-resonances of the ring\cite{deo_stub}.  In all the figures we
have taken dimensionless parameters, the current densities $J_u$ and
$J_d$ plotted in the figures are in their dimensionless form given by
dividing $J_u$ and $J_d$ by $\frac{e\hbar k}{m}$.

\section{Evanescent AB interferometry in presence of resonant states}

\begin{figure}[h]
\protect\centerline{\epsfxsize=3.0in\epsfbox{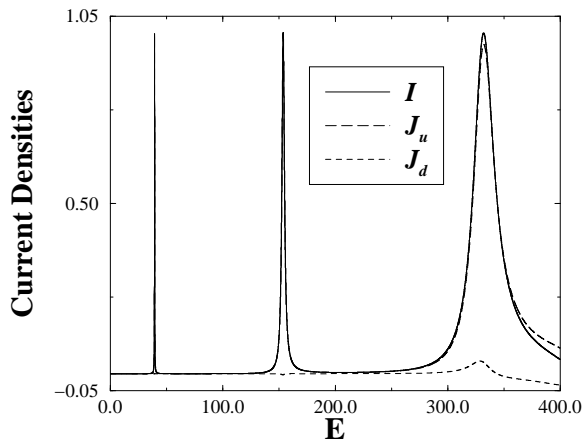}}
\caption{Current densities for  zero field and increasing
  Energy. Current magnification seen at $E \le V$. Barrier potential
  $V=400.0$. Lengths are $lu1 = lu2 = 0.1$, $lw=0.4$ and $ld=0.4$.}
  \end{figure}

\begin{figure*}
  \protect\centerline{\epsfxsize=5.5in\epsfbox{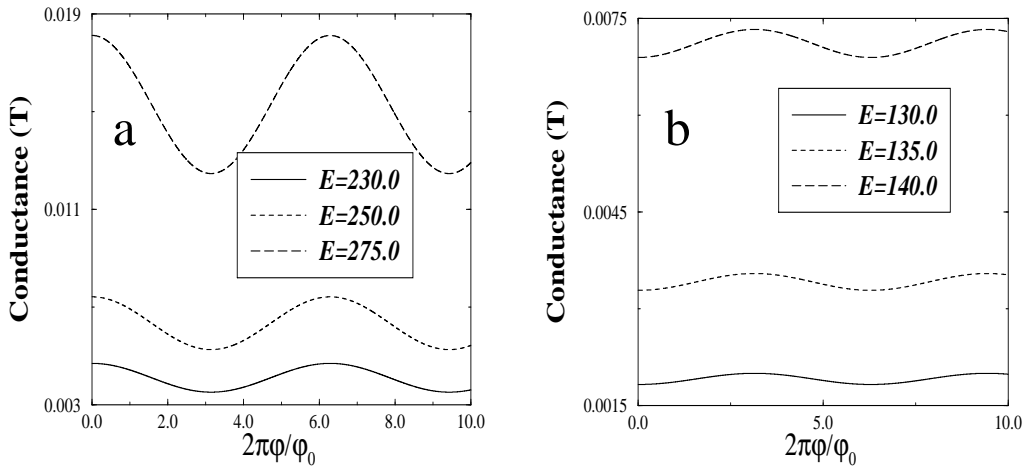}}
  \caption{Conductance for increasing
    field. (a) For Fermi energies to right of second resonance in FIG.~5. Same
    physical parameters as in FIG.~5. (b) For Fermi energies to left
    of second resonance in FIG.~5. Same physical parameters as in FIG.~5.}
\end{figure*}

Having established that evanescent mode consideration shows
systematics in MC but absence of current magnification while transport
in the propagating regime shows opposite behavior, no systematics in
MC but current magnification is evident. Now we turn our attention to
the problem as we had formulated it in our second configuration, i.e.,
evanescent mode transport in an Aharonov-Bohm ring with a quantum well
in it's upper arm to simulate resonances. The need to incorporate
resonant states arises from the fact that presence of unavoidable
defects and other localized resonant states in such small devices can
mar switching action and hamper device performance\cite{switch}. To
analyze this if a well sustaining a few resonant states can be
incorporated in the Aharonov-Bohm interferometer, then device
performance can be tested.  To this end we justify the inclusion of a
quantum well in one of the arms. The parameters defining the quantum
well are mentioned in the caption of FIG.~5, in dimensionless form,
$l_{w}$ represents width of the quantum well, while $V$ defines the
potential barrier's, in the well region the particle is assumed to be
free

In FIG.~5, we plot the output current density and the current
densities in upper and lower arms as a function of the Fermi energy
for asymmetric arm lengths (as in FIG.~1). We encompass three resonant
states of the well. The length and height of barrier potential are
mentioned in the figure caption. We assume the particle to be free in
the well, i.e., in the well, potential is zero. The resonances in
FIG.~5 are all of Breit-Wigner type in contrast to transmission
resonances for propagating mode transport which can be either of
Breit-Wigner or Fano type\cite{deo_stub}. In contrast to evanescent
transport without a quantum well, at Fermi energies comparable to
barrier potential but still in the evanescent regime, current
magnification is seen. The current density($J_{u}$) of the upper arm
is more than output current while that of lower arm($J_{d}$) is
negative. In this regime a circulating current continually flows in
the ring even for evanescent wave transport in the region outside the
well. We have also seen separately that if potential well supports
many resonant states(second configuration) then also current
magnification occurs at Fermi energies comparable to barrier potential
and {\it interestingly} it is always to the right of the last
resonance (E$\le$V).

To see the effect of resonant states on the MC we plot conductance for
different Fermi energies on both sides of the second resonant peak of
FIG.~5, in FIG.~6(a) and FIG.~6(b). In FIG.~6(a) we plot the
magneto-conductance as a function of flux for Fermi energies to right
of the second resonance and left of third resonance as in FIG.~5. In
FIG.~6(a) we see that the small field differential magneto-conductance
is still negative and hence systematic behavior can still be seen,
while in FIG.~6(b) we plot the magneto-conductance in the energy
window in-between first and second resonances, and see that the small
field differential magneto-conductance is positive. Thus we can draw
some uniform conclusions as to where systematic action is to be seen.
In particular nature of slope of magneto-conductance is always
negative for Fermi energies to the left of first resonance, and then
the nature of this slope (being negative or positive) alternates as we
cross each successive resonance. This systematic is possible due to
first, a gradual change in the transmission phase by $\pi$ across the
well while crossing the resonance and second due to the absence of
transmission zeros in this system. The transmission zeros can lead to
abrupt change in phase by $\pi$ of the
conductance\cite{imry,fano_deo}. This behavior is unlike that shown in
case of transport via propagating modes as seen in FIG.~2(b), where
the systematics in MC cannot even be predicted due to presence of
transmission zeros.

\section{Question of Fano resonances}

\begin{figure*}
\protect\centerline{\epsfxsize=5.5in\epsfbox{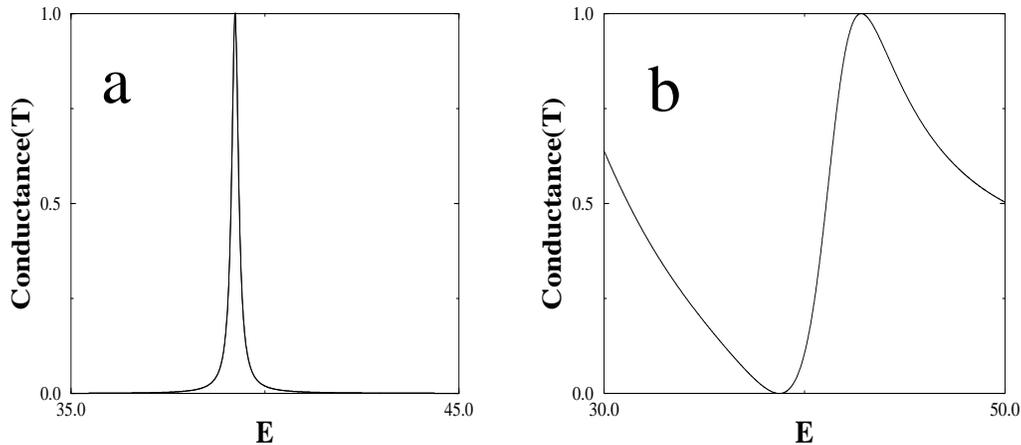}}
\caption{(a) Evanescent transport for zero field and increasing
  energy. Barrier potential $V=400.0$ and lengths are
  $lu1=lu2=0.1,lw=0.4$ and $ld=0.4$, all in dimensionless units. (b)
  Fano resonant line-shapes. Lower arm is free (barrier absent). Same
  physical parameters as in (a).}
\end{figure*}

Recently Fano resonances have been observed in a mesoscopic AB ring
with a quantum dot embedded in it's upper arm\cite{kobay}. Fano
resonances arise when a discrete set of states are coupled to the
continuum\cite{kobay}.  Especially Fano resonances are caused by
interference of two alternative paths\cite{imry}, a resonant and a
non-resonant one. In the geometry considered a resonant path is via a
quantum well (having resonant energy levels) and a direct non-resonant
path is via the lower arm of the ring (FIG.~1).  Transmission
resonances are generally of two types Breit-Wigner and Fano. Fano
line-shapes are asymmetric as opposed to the symmetric Breit-Wigner
line-shapes. In contrast to Breit-Wigner forms, Fano resonances are
charecterized by  zero-pole
pair structure in the complex energy plane of the scattering
amplitude\cite{porod,bagwell}.

In FIG.~7(a) we have plotted the conductance (in effect the
transmission coefficient in dimensionless form) as function of the
Fermi energy.  We see resonant transport around the quasi bound states
of the well (around $E=39.0$).  The line shape is that of Breit-Wigner form.
We do not observe any Fano line-shapes as long as $E<V$. As mentioned
earlier Fano line-shapes arise from the interference of two
alternative paths, one resonant and another direct. Our results
clearly indicate when the transport across the direct path is via
evanescent modes, then Fano line-shapes are absent. To recover Fano
line-shapes, direct path should be a propagating one. To see this
explicitly we set potential in the lower arm to be zero (configuration
three). So that electron traverses the lower arm as a propagating
wave. For this case we have plotted in FIG.~7(b), the conductance
versus Fermi energy for the same physical parameters as in FIG.~7(a).
We clearly see the Fano line-shapes, i.e., line-shape is asymmetric
and transmission exhibits zero around the same Fermi
energy interval. The results presented in FIG.'s~7(a) and 7(b) are in
the absence of magnetic field. Increasing the magnetic field which
amounts to breaking the existing time reversal symmetry will lead to
lifting of zeroes except when the flux piercing the loop is an
integral or half integral multiple of the flux quantum\cite{broken}.

\section{Conclusions}

To conclude, when electron traverses the entire circumference of the
ring as an evanescent wave, the current magnification effect is
absent. In this regime small field differential magneto-conductance is
always negative. Inclusion of a well in an otherwise evanescent
transport supports current magnification. The magnetoconductance shows
systematic behavior even in this regime. The initial slope of the
magnetoconductance alternates as we cross successive resonance peaks.
The transmission resonances one observes in this set up (resonant
states in an evanescent AB interferometer) are of Breit-Wigner type in
contrast to the Fano line shapes expected because of the coupling
between a resonant and non-resonant path. One recovers Fano resonances
if transport through non-resonating path is via propagating modes.
This fact emphasizes the need of a direct propagating path for
observance of Fano line shapes in transmission.


\begin{thebibliography}{99}

\bibitem{steinberg} A. M. Steinberg, preprint cond-mat/9710046.

\bibitem{webb} S. Washburn and R. Webb, Rep. Prog. Phys. { \bf 55}, 1311 
  (1992).   
  
\bibitem{levy} L. P. Levy, G. Dolan, J. Dunsmuir and H. Bouchiat, \prl
  {\bf 64}, 2074 (1990); D. Mailly, C. Chapelier and A. Benoit, \prl
  {\bf 70}, 2020 (1993).

\bibitem{gia}Y. Gefen, Y. Imry, and M. Ya. Azbel, \prl { \bf 52}, 129
  (1984).
  
\bibitem{kobay} K. Kobayashi, H. Aikawa, S. Katsumoto and Y. Iye, \prl
  { \bf 88}, 256806 (2002);U. Fano, Phys. Rev. { \bf 124}, 1866
  (1961);A. A. Clerk, X. Waintal and P. W. Brouwer, \prl { \bf 86},
  4636 (2001).
  
\bibitem{colin_cm} Colin Benjamin and A. M. Jayannavar, \prb
  { \bf 64}, 233406 (2001).

\bibitem{deo_stub} P. S. Deo and A. M. Jayannavar, \prb { \bf 50},
  11629 (1994); T. P. Pareek, P. S. Deo and A. M. Jayannavar, \prb
  { \bf 52}, 14657 (1995).

\bibitem{switch} P. S. Deo and A. M. Jayannavar, \mpl { \bf 8},
  301 (1994); B. C. Gupta, P. S. Deo and A. M.  Jayannavar , \ijmp{ \bf
    10}, 3595 (1996).



\bibitem{coupled} T. P. Pareek and A. M. Jayannavar, \prb
  { \bf 54}, 6376 (1996); A. M. Jayannavar and P. SinghaDeo, \prb { \bf 49},
  13685 (1994).

\bibitem{xia} J-B. Xia, \prb { \bf 45}, 3593 (1992).

\bibitem{maple} MAPLE V version 5, www.maplesoft.com.



\bibitem{colin_de} Colin Benjamin and A. M. Jayannavar, \prb
  { \bf 65}, 153309 (2002), and references therein.

\bibitem{mosk} M. V. Moskalets, Euro. Phys. Lett. { \bf 41}, 189
  (1998).

\bibitem{choi} T. Choi, C. M. Ryu and A. M. Jayannavar, \ijmp { \bf
    12}, 2091 (1998); T. Choi, C. M. Ryu and A. M. Jayannavar,
  preprint cond-mat/9808245.

\bibitem{imry} Ora-Entin Wohlman, A. Aharony, Y. Imry and Y. Levinson, 
  \ltp { \bf 10}, 3595 (2001), preprint cond-mat/0109328. 

\bibitem {fano_deo}  P.S. Deo and A.M. Jayannavar,
  Mod. Phys. Lett. B { \bf 10}, 787 (1996). 

\bibitem {porod} W. Porod, Z. Shao and C. S. Lent,  \prb { \bf 48}, 8495 (1993).
\bibitem {bagwell} E. Tekman and P. F. Bagwell,  \prb { \bf 48}, 2553 (1993).

\bibitem{broken} T. S. Kim, S. Y. Cho, C. K. Kim and C. M. Ryu, \prb { \bf
    65}, 245307 (2002).
  

\end{thebibliography}
\end{document}